\documentclass[runningheads]{llncs}
\usepackage{subcaption}
\usepackage{graphicx}

\usepackage{listings}
\usepackage[usenames,dvipsnames]{xcolor}
\definecolor{lightyellow}{RGB}{255, 248, 225}
\lstdefinelanguage{SimPy}{%
  language     = Python,
  morekeywords = {yield},
}
\usepackage{hyperref}

\begin{document}
\title{A Python-based Mixed Discrete-Continuous Simulation Framework for Digital Twins}

\titlerunning{Mixed Discrete-Continuous Simulation Framework for Digital Twins}
%
\author{Neha Karanjkar\inst{1}\orcidID{0000-0003-3111-1435} \and Subodh M. Joshi\inst{2}\orcidID{0000-0002-9239-8866}}
\authorrunning{N. Karanjkar and S. M. Joshi}
%
\institute{Indian Institute of Technology Goa, India \\
\email{nehak@iitgoa.ac.in}\\
\and
Department of Computational and Data Sciences, Indian Institute of Science, Bangalore, India\\
\email{subodhmadhav@iisc.ac.in}}
\maketitle              
%
\begin{abstract}
	The use of Digital Twins is set to transform the manufacturing sector by
	aiding monitoring and real-time decision making. 
	For several applications in this sector, the system to be modeled consists
	of a mix of discrete-event and continuous processes interacting with each
	other.
	Building simulation-based Digital Twins of such systems necessitates
	an open, flexible simulation framework which can support easy modeling
	and fast simulation of both continuous and discrete-event components, and
	their interactions.

	In this paper, we present an outline and key design aspects of a
	Python-based framework for performing mixed discrete-continuous
	simulations.  The continuous processes in the system are assumed to be
	loosely coupled to other components via pre-defined events. For example, a
	continuous state variable crossing a threshold may trigger an external
	event. Similarly, external events may lead to a sudden change in the
	trajectory, state value or boundary conditions in a continuous process.
	We first present a systematic events-based interface using which such
	interactions can be modeled and simulated. We then discuss implementation
	details of the framework along with a detailed example. In our implementation,
	the advancement of time is controlled and performed using the event-stepped 
	engine of SimPy (a popular discrete-event simulation library in Python). The
	continuous processes are modelled using existing frameworks with a
	Python wrapper providing the events interface.  We discuss possible
	improvements to the time advancement scheme, a roadmap and use cases for the
	framework.

\end{abstract}
	\section{Introduction} 

  A Digital Twin refers
	to a digital representation (computer model) of a real system that is
	continuously kept in sync with the system using periodic sensing of
	its health parameters and is used for prediction,
	optimization and control of the real system.  
  The use of Digital Twins, aided by advancements in 
  Internet of Things (IoT) technologies and  Machine Learning (ML) based 
  analytics is set to   transform many sectors such as 
	manufacturing, healthcare, urban planning, energy and transportation.
  While a data-driven model may suffice as a Digital Twin for some applications,
  a detailed simulation model is often used to create Digital Twins
  of complex processes in the manufacturing sector.
  Such systems often consist of a mix of discrete-event processes and 
  continuous processes interacting with each other. 
  For example, in \cite{HUDA2002} the modeling of food processing systems is described 
  which requires a detailed
  simulation of continuous phenomena as well as discrete event processes
  with inter-dependencies.

  This paper presents the key design aspects and implementation details 
  for a Python-based Mixed Discrete-Continuous Simulation (MDCS)
  framework targeted for creating Digital Twins.
  A preliminary outline and motivation for this framework was first presented in \cite{KaranjkarJoshi2021}.
  This paper expands on the implementation aspects and presents a detailed 
  example to demonstrate the event-based interface used 
  for incorporating and simulating continuous processes in this framework.

  In this section, we first present a broad overview of existing simulation approaches and frameworks
  used for building Digital Twins.
  We then summarize the motivation and design goals for a Python-based MDCS framework.
  In Section \ref{sec:framework} we present the details of the framework and 
  describe the events-based interface that can be used for integrating continuous
  processes into a discrete event simulation engine. 
  In Section \ref{sec:modeling_example}, we present a detailed example that illustrates
  the key aspects and use of the framework. 
  While one of the continuous entities (a fluid tank) used in the example is similar 
  to that described in \cite{KaranjkarJoshi2021}, the example also incorporates a continuous
  process describing heat dissipation in a two-dimensional (2D) hot-plate that 
  interacts with the fluid tank and other discrete event processes in the system.
  Unlike the fluid tank (where the state-updates are modeled by simple, linear equations),
  the simulation of the transient heat transfer in the hot-plate requires resolution of
  the governing Partial Differential Equations (PDEs) using a Finite Difference Method (FDM).
  We present detailed simulation results for a test case where both types of 
  components interact with each other.
  An improved scheme for time-advancement and a roadmap for further development 
  are discussed in Sections \ref{sec:timeAdvancement} 
  and \ref{sec:futureWork} respectively.

\subsection{A Review of Simulation Approaches for Digital Twins} 
  The design of a simulation framework for
	Digital Twins is driven by the characteristics of the system to be modeled.
  The methodologies and approaches for numerical simulations depend 
  on the type of the system under consideration, 
  that is, whether the system is continuous, discrete or 
  contains a mix of both types of entities.
  While some system models may necessitate a continuous simulation framework
	\cite{Aversano2020,Molinaro2021}, a discrete-event simulation might
	suffice for other kinds of Digital Twins \cite{Agalianos2020a}.
  A summary of challenges and
	desired capabilities associated with the simulation of Digital Twins is
	presented in \cite{Shao2019}.  
  In the context of Digital Twins, 
	continuous systems are the ones subjected to continuous evolution of
	the state variables in time.  
  A few examples of such systems include transient transfer of
  heat, unsteady fluid flows such as air-flow over a wind turbine,
  chemical reactions etc.  
  The state variables for such systems often undergo a continuous change in 
  time (unless the system reaches a steady state). 
  Accurate simulation of these systems necessitates solving
  the governing equations, which often take the form of 
	ordinary/partial differential equations (O/PDE) or mixed
	differential-algebraic equations (DAE).  
  This in turn requires using appropriate numerical schemes, for example 
  Finite Volume, Finite Element or Finite Difference Methods as well as 
  schemes for advancing the solution in time. 
  The time-stepping methods can be explicit, implicit or mixed type
  such as the IMplicit-EXplicit (IMEX) method.
  In most of the time-stepping methods, the
  time step-size may be either fixed, or adjusted
	dynamically over the course of a simulation.  A detailed description of
	continuous processes and their simulation aspects can be found in
	\cite{Leveque90,Cellier06} and the references therein.  In practice,
	frameworks such as FEniCS \cite{dolfin10}, Deal II \cite{dealIIa}, OpenFOAM
	\cite{weller98} are used for continuous multiphysics simulations of complex
	systems.  Numerical techniques such as reduced order models (ROM)
	\cite{chinesta11,feng05} and Machine-Learning (ML) based metamodels
	\cite{simpson01} may be used instead of the high fidelity models to reduce
	the overall computing cost.  In the recent times, Machine Learning is
  increasingly being used for scientific computing 
  \cite{Aimone2017,Karniadakis2021,Brunton2020}.
  In this paper, we use the classical Finite Difference Method (FDM)
  for simulation of the continuous entities, however, we plan to 
  explore the other simulation frameworks including Machine Learning in
  future work.

	Unlike the continuous processes described earlier,
  discrete processes are characterized by changes in the state of
	the system occurring only at discrete (countable) instants of time,
	referred to as {\em events}.  Discrete-event simulation is prominently divided
	into two approaches, viz. event-stepped and cycle-stepped approaches.
	We refer \cite{Hill2007} for a detailed description of the approaches used for
  simulation of discrete event systems.  
  Discrete Event System Specification
	(DEVS) and a subsequent generalization (GDEVS) are the two main formalisms 
  used for specification and simulation of discrete-event systems
	\cite{Zeigler1989b,Giambiasi2001}.  Agalianos et. al. present an overview
	of issues and challenges for discrete-event simulation in the context of
	Digital Twins \cite{Agalianos2020a}.  There exist several proprietary as
	well as open source libraries and softwares for discrete-event simulations.
	A review of open source discrete simulation softwares is presented in
	\cite{dagkakis2016}.

	Systems containing both discrete-event and continuous processes
  are termed as Mixed Discrete-Continuous (MDC) systems. 
  Simulation of MDC systems is particularly challenging since it 
  requires to satisfy the constraints imposed by both the continuous 
  and the discrete entities involved.
  Different methods and techniques have been proposed in literature 
  for simulation of MDC systems.
  A quantization based integration method was proposed by Kofman et. al. 
  for simulation of hybrid systems \cite{Kofman2004}.  
  Nutaro et. al. propose a split system approach in which
	a-priori knowledge about the discrete-continuous structural split in the
	model can be used for performing efficient simulation \cite{Nutaro2012}.
	Klingener describes approaches that can be used to get a non-modular
	simulation framework \cite{Klingener1995,Klingener1996}.
	An approach called Discrete Rate Simulation has been proposed in 
  \cite{Damiron2008} for
	simulating linear continuous models (such as constant-rate fluid flows)
	within a discrete-event framework.
  A usecase for this approach has be demonstrated by Bechard et. al. in \cite{Bechard2013}.
	Eldabi et. al. present a detailed overview of various strategies used for 
  MDC simulation in	\cite{Eldabi2019}.

\subsection{Motivation and Design Goals}

	While the ability to perform mixed discrete-continuous simulations
	is a key requirement for the development of the framework presented
	in this paper, the other design goals are as follows:
	
	\begin{enumerate} 
	
	\item The ability to model heterogeneous
	systems containing different kinds of continuous processes, each possibly
	requiring a different numerical method for its solution and/or different
	characteristic time-step sizes. 
	
	\item Support for capturing the effect of
	periodic sensor updates from the real system on the model's state. 
	
	\item The ability to perform real-time simulation. 
	
	\item The framework should be open-source and flexible. It should be easy to integrate existing libraries
	for enabling analytics and visualization  (e.g. optimization, machine
	learning, data handling, scientific computing and plotting libraries) into
	the framework.  
	
	\item The language used by the framework should support
	modular descriptions and the use of object oriented features for modeling
	complex systems with many interconnected components.  
	
	\end{enumerate} 
	
	While there exist a number of frameworks that are targeted separately for either
	continuous simulation or discrete-event simulations, the requirement of
	simulating both discrete and continuous processes together, possibly
	interacting with each other, introduces some challenges. A majority of
	the existing frameworks for MDC simulations are either commercial or
	domain-specific, with exceptions such as OpenModelica\cite{openmodelica20}.
	However, there still exists the need for a mixed simulation framework that
	is written in a general-purpose, object-oriented language which allows
	integration with existing continuous simulation frameworks.  Python is an
	attractive choice for implementing such a framework because of its wide
	user base, ease of use and the availability of a number of libraries for
	analytics and visualization.

\section{Framework for Mixed Discrete-Continuous Simulation}\label{sec:framework}

	In this section, we present the basic definitions and assumptions,
	describe the main aspects of the framework (including the events interface) 
	and discuss the implementation details.
	The system to be modeled can be thought of as a mix of continuous and discrete
	entities that interact with each other.
	\begin{itemize}
	\item An {\bf entity} in this context is a collection of state variables, methods and
	processes representing a particular object to be modeled in the system. 
	
	\item A {\bf discrete entity} refers to a process whose state can change only at
	discrete time instants (events). 
	
	\item A {\bf continuous entity} is an entity whose state may be considered to change
	continuously with time and may require continuous simulation/monitoring. 
	\end{itemize}
	When simulating discrete and continuous processes interacting with each other,
	the fundamental questions related to the advancement of time have been 
	addressed by formalisms proposed for hybrid simulations, for e.g.  \cite{Nutaro2012}.  
	However, from an implementation perspective, the simulation approaches 
	can be broadly classified into two categories as
	summarized below:

		\begin{enumerate} 
		\item[{\textbf{(A)}}] In the first approach, the
		advancement of time is controlled by a single time-stepped 
		continuous simulation loop.  The discrete events to be modeled are embedded into the continuous
		simulation code as conditional updates to the state variables or boundary
		conditions during simulation. These updates may occur at pre-defined
		time-steps  or whenever a certain  condition on the state variables is met
		(for example, when the value of the state variable crosses a particular
		threshold).  The step-size for advancing time is fixed and determined by the
		stability considerations of the numerical scheme used for continuous
		simulation. 
		If the required step-size differs across multiple continuous entities in
		the system, the smallest of the step-sizes needs to be used for updates
		in {\em all} of the continuous entities.
		Such an approach is well suited for systems mainly consisting of tightly
		coupled continuous entities.\\

	  \item[{\textbf{(B)}}] In the second approach, time advancement is handled by
	  an event-stepped discrete-event simulation algorithm.  Simulation of 
	  continuous entities is performed by invoking their state-update functions 
	  periodically or at selected time-instants from within the event-stepped loop.
	  Interactions of a continuous entity with other entities in the system are
	  modeled via {\bf events}. The continuous entity must generate and advertise events 
	  when certain conditions on its state variables are met (for example, when the 
	  state value crosses a certain threshold), so that external activities can be triggered
	  on the occurrence of this event. Similarly, external events may cause a sudden change in the
	  state values or trajectory of the continuous entity. Therefore, whenever such external
	  events occur, there must be a mechanism to update the state/boundary conditions of the continuous
	  entity and invoke its state update function.
	  This approach is well-suited when the continuous
	  entities in the system are few and loosely coupled.  
	  \end{enumerate}

	  Approach {\bf(B)} is particularly suited for simulation of Digital Twins in
	  manufacturing and process engineering domains       since the systems to be
	  modeled are heterogeneous, typically consisting of a larger number of
	  discrete entities and a few continuous entities that are loosely coupled and
	  interact in well-defined ways.  From a modeling perspective, 
	  describing interactions between components through events 
	  can lead to modular descriptions that are easier to read, maintain and debug.
	  We present a framework based on this approach and 
	  describe how the interactions can be modeled and time advancement can be performed.

	\subsection{The Events Interface} \label{sec:eventsInterface}
\begin{figure}[] 
	\centering
	\includegraphics[width=0.6\textwidth]{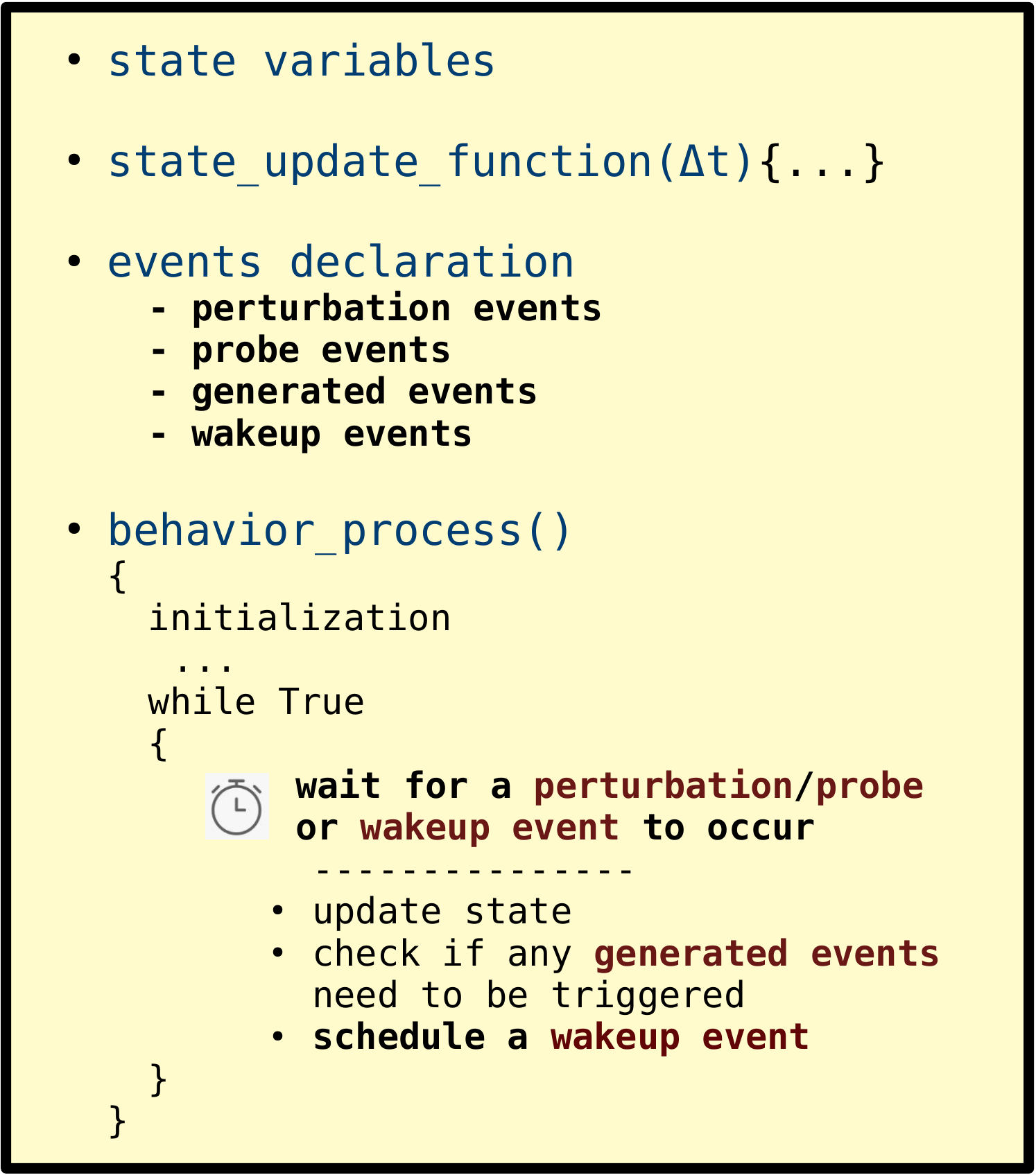}
	\caption{Wrapper for a continuous entity with the events serving as its interface\cite{KaranjkarJoshi2021}} \label{fig:continuous_entity}
	\end{figure}

	In our framework, time advancement is performed using the 
	event-stepped algorithm of a general purpose discrete-event
	simulation framework (such as SimPy). To incorporate continuous entities
	into the simulation, a wrapper for each continuous entity is made, which provides an 
	interface as summarized in Figure \ref{fig:continuous_entity}.
	A continuous entity is characterized by its state variables and a
	state-update function. The modeler also needs to provide a definition of
	all events that may be generated by the entity and affect the external
	world and vice-versa. These events can be classified into four types as
	follows:
	\begin{enumerate} 
	
	\item \textbf{Perturbation event: } An externally-created event that may affect
	the state/trajectory of the continuous entity.

	\item \textbf{Probe event: } An externally-created event which involves querying the
	state of the continuous entity and thus necessitates updating its state up-to a
	given time.  
	
	\item \textbf{Generated event: } An event triggered by the continuous
	entity itself as a consequence of its state update which may affect other
	entities in the system.  
	
	\item \textbf{Wakeup event: } An event scheduled by
	the continuous entity itself for performing state updates after a fixed time
	step or for creating generated events whose time of occurrence can be predicted in
	advance.  
	
	\end{enumerate}
	The behaviour of the continuous entity is modeled as a discrete-event process.  
	This process is activated whenever a perturbation, probe or a
	wakeup event for the entity occurs.  When activated:
	\begin{enumerate} 
	\item
	The state updates for the entity up-to the current time are computed.  
	
	\item If any condition for generating output events is met (for example, the state
	variable crossed a threshold), the generated events are triggered.
	
	\item The continuous entity schedules a wakeup event for itself after
	a particular time interval.  This time interval is determined based on one of two 
	approaches as follows:

		\begin{itemize} 
		
		\item {\bf (a) Predictive time-stepping:} If the current trajectory of the state values in the entity
		is known, and if all of the output events can be predicted ahead of
		time based on this trajectory, the wakeup can simply be scheduled at
		the time the earliest output event is predicted to occur.  
		(This is illustrated via the example of a fluid tank described in Section \ref{sec:tank}.

		\item {\bf (b) Fixed time-stepping:} If the exact time instants of events that will be generated by this entity
		cannot be predicted ahead of time, the state needs to be updated after regular 
		time intervals that are small enough, by scheduling a wakeup event periodically. 
		The step-size may need to be chosen based on numerical stability requirements.  
		(This is illustrated via the example of a heater entity described in Section \ref{sec:heater}.
		
		\end{itemize}
	
	\end{enumerate}
	At each iteration of the event-stepped algorithm, the simulation time
	is advanced to the time-stamp of the earliest scheduled event in the global
	event list.  All events scheduled to occur at this time are executed and the
	processes waiting on this event are automatically triggered (using a mechanism 
	such as callbacks) provided by the discrete-event framework.

	It is to be noted that if the system has multiple continuous entities 
	requiring fixed time-stepping, and if the chosen step-sizes of these entities differ, 
	then in this approach, each entity will be woken up periodically 
	as per its own step-size only, unless there is an 
	external event affecting it. This leads to an efficient simulation, where
	some entities may be updated with a coarse time-step while some may use a 
	finer time-step.

	\subsection{Implementation} \label{sec:implementation}

	We have implemented the MDCS framework using SimPy\cite{SimPY20}, a discrete-event simulation library in Python.
	Processes in SimPy are implemented using Python's generator functions and can be used to model
	active components. The processes are managed by an {\it environment} class,
	which performs time advancement in an event-stepped manner using a global
	event queue.  The system to be modeled can be described in Python using a
	few SimPy constructs and does not require the user to learn a new modeling
	language. SimPy also supports real-time simulation.  However, SimPy is
	designed for discrete-event simulation and currently offers no features for
	modeling continuous systems \cite{SimPY20}. The events-based interface presented
	in this paper can be implemented as an abstract wrapper class to allow 
	integration of continuous entities.
	The modeler needs to create the wrapper class for each unique type of continuous entity
	describing the aspects summarized in Figure \ref{fig:continuous_entity}.
	When the simulation of the continuous entity needs to be performed using an external solver 
	(such as Deal II or OpenFOAM), the state update function in the abstract class
	serves as a wrapper for invoking the state updates via the external continuous solver.
	Similarly, the continuous solver code needs to be instrumented to detect conditions
	that must trigger events, and set a flag variable in the wrapper. State or boundary 
	value updates to the continuous entity can be implemented via global shared variables that may be updated
	by external processes, but are read at each iteration in the continuous solver code.
	The implementation is straightforward if the simulation and state updates for the continuous entity
	are also described as Python code. In the example presented in Section \ref{sec:modeling_example}  the 
	state updates of continuous entities are performed directly by including their description
	as Python modules.

\section{A Modeling Example} \label{sec:modeling_example}

	We present a modeling example to illustrate the key aspects of the MDCS
	framework.
	The system to be modeled consists of discrete-event processes as well as
	two types of entities that are modeled in the continuous domain. The first
	type of entity (a fluid tank) has simple, linear state-update equations.
	Simulation of this entity can be performed using the predicted
	time-stepping approach. The second entity is a heater with a square-shaped
	plate which can be heated from two opposite sides. 
	The temperature within the plate
	varies continuously with time and also across the length and breadth of the
	plate. The diffusion of heat in the plate is simulated using fixed
	time-stepping approach, with the time-step size determined by numerical
	stability requirements.  The complete system to be simulated consists of
	both types of components interacting with each other.

	We first describe each of these components and their simulation approaches
	in detail.  We then describe the complete system consisting of the fluid
	tank and the heater instances interacting with each other as well as with
	other discrete-event processes.  We show how the interactions can be
	described in the framework, and present detailed simulation results.

\subsection{Heater}\label{sec:heater}

\subsubsection{Description: }
\begin{figure}
\centering
\includegraphics[width=\textwidth]{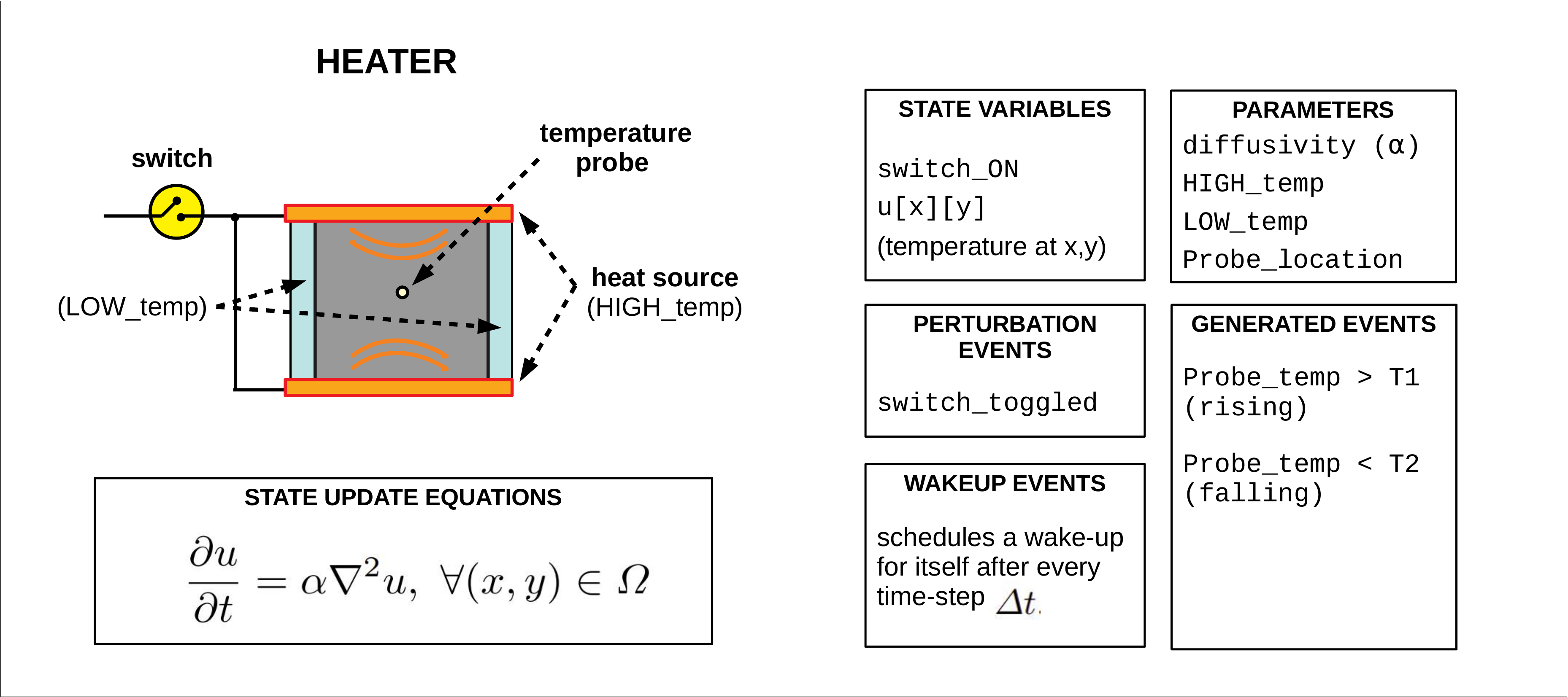}
\caption{Description of the Heater model} \label{fig:heater}
\end{figure}

The physical system consists of a square-shaped hot-plate with each side $1$m long
and made of a composite material.
Two opposite edges of this plate can be heated with the help of heating coils. 
We make the simplifying assumption that the heating coils instantly 
reach the steady-state high temperature value without any lag when switched on. 
The remaining two edges of the hot-plate are maintained at a constant lower temperature at all times. 
After the heating coils are turned on, 
heat dissipates in the hot-plate and eventually the temperature distribution attains 
a steady state profile. How quickly the heat dissipates depends on the thermal 
conductivity, the specific heat capacity and the density of the material of the hot-plate. 
A physical parameter called as thermal diffusivity ($\alpha$, units m$^2$/s) is often used to describe 
the resistance offered by any material for heat dissipation. It takes into account the collective
effects due to all the material properties mentioned above. 
We assume the hot-plate to be made of a highly conductive composite material 
with thermal diffusivity equal to $0.0005$m$^2$/s. 
When the heater is turned off, the heat starts flowing out via all the four edges.
We assume the convective and the radiative losses to be negligible and the
entire heating and cooling of the hot-plate takes place only through the boundaries. 
We measure the temperature of the hot-plate with the help of
a temperature probe. 
The framework allows specifying the location of the probe at any location. 
We assume the probe to locate at the center of the plate in this example. 
For this test case, we specify the heater temperature (when turned {\tt ON}) as $100^0$C
({\tt HIGH\_temp}) and the
room temperature, which is also the temperature maintained constantly at the other two edges, as
$25^0$C ({\tt LOW\_temp}). 
Further, when the heating is switched {\tt OFF}, all the four boundaries are assumed to be at the uniform 
temperature equal to {\tt LOW\_temp}.
Figure \ref{fig:heater} shows a schematic of the hot-plate along with the state-variables and 
the state-update equations governing the conductive heat transfer. 

\subsubsection{Numerical Simulation: }
Let the hot-plate be modeled by a square shaped computational domain $\Omega := [0,1] \times [0,1]$.
Let the temperature in the plate be described by $u(x,y)$, where $(x,y)$ is any physical location. 
Further, time is denoted by $t$.
The thermal diffusivity $\alpha$ is a material-specific property, which depends on the
thermal conductivity of the material, its specific heat capacity and its density. 
The heat conduction through a homogeneous material is described mathematically by the 
following Partial Differential Equation (PDE).
\begin{equation}\label{Eq:heater}
    \frac{\partial u}{\partial t} = \alpha \nabla^2 u, ~ \forall (x,y) \in \Omega
\end{equation}
The boundary conditions are as follows.
    $u(x,0) = u(x,1) =$ Heater temperature: i.e. {\tt HIGH\_temp} (heater {\tt ON}) 
                                              or {\tt LOW\_temp} (heater {\tt OFF}) and 
    $u(0,y) = u(1,y) =$ {\tt LOW\_temp} $\forall$ t. 

To solve this equation numerically, we use the Finite Difference Method (FDM).
First, the domain $\Omega$ is discretized in a structured, uniform grid with the grid-spacing $h$.
Let the number of grid-points in each direction $x$ and $y$ be $N$, i.e. $N^2$ in total.  
Therefore, the location of any grid-point $(i,j)$ is given as $(x_i, y_j) = (ih, jh), ~ i,j=0,1,2,...,N-1$ . 
Similarly, let the time-step for advancing the numerical simulation be $\Delta t$.
The temperature at any grid-point $(i,j)$ at time level $k$ is denoted as $u_{i,j}^k$.
The difference equation corresponding to the governing equation (\ref{Eq:heater}) can be written
as (derived from Taylor's series expansion, i.e. standard FDM formulation):
\begin{equation}\label{Eq:heaterDifference}
    \frac{u_{i,j}^{k+1} - u_{i,j}^{k}}{\Delta t} = \alpha 
                                                 \left( 
                                                 \frac{u_{i+1,j}^{k} - 2u_{i,j}^{k} + u_{i-1,j}^{k}}{h^2}
                                                +\frac{u_{i,j+1}^{k} - 2u_{i,j}^{k} + u_{i,j+1}^{k}}{h^2} 
                                                 \right)
\end{equation}
with the Dirichlet boundary conditions given as
$$
    u_{0,j}^k = u_{N-1,j}^k = \textnormal{{\tt LOW\_temp}}
$$
and
$$
    u_{i,0}^k = u_{i,N-1}^k = \textnormal{{\tt HIGH\_temp} (heater {\tt ON}) or {\tt LOW\_temp} (heater {\tt OFF})}
$$
Equation (\ref{Eq:heaterDifference}) can be further simplified as
\begin{equation}\label{Eq:heaterDifference1}
    u_{i,j}^{k+1} = \gamma 
                                                 \left( 
                                                 {u_{i+1,j}^{k} + u_{i-1,j}^{k}}
                                                +{u_{i,j+1}^{k} + u_{i,j+1}^{k}}
                                                -{4u_{i,j}^{k}}  
                                                 \right)  
                                                 + u_{i,j}^{k}
\end{equation}
where, $\gamma = \alpha \Delta t / h^2$.
In the beginning of the simulation, the initial conditions $u_{i,j}^0 = \textnormal{\tt LOW\_temp}, \forall i,j$
hold true for the entire domain. Equation (\ref{Eq:heaterDifference1}) is solved iteratively to 
advance the simulation in time. 
This time-stepping scheme is also known as the Forward-Euler (FE) explicit timestepping scheme.
It is to be noted that, the FE timestepping scheme is only conditionally stable, i.e. the
time-step value needs to be `small' enough to yield stable computations. 
According to the Von-Neumann stability analysis, the time-step
value comes out to be $\Delta t \le h^2/(4\alpha)$.
i.e. this is the largest value of $\Delta t$ that can be safely 
used for advancing the solution in time.
This in turn also dictates the frequency of wake-up events that needs to be scheduled. 
An efficient, vectorized implementation of Equation  (\ref{Eq:heaterDifference1})
is performed using Numpy, a Python library for scientific computing. 
Python also makes it convenient to incorporate the numerical implementation of 
the heater in the discrete event framework.  

\subsubsection{Interface with the Discrete-Event Simulation Framework}
The heater entity is implemented as a Python class.
The state variables, parameters and the state-update
function (i.e. the numerical solution of the 
governing PDE given by Equation \ref{Eq:heater}) 
become the members of this class.
The interaction of the heater with the environment
happens via the following types of events:

\begin{enumerate}
  \item {\bf Perturbation Events:} These are the
    events which can cause the heating coils to turn on or off. 
    This in turn results in heat transfer 
    into or out of the system affecting the temperature distribution. 
  \item {\bf Probe Events:} We can specify the location of the probe on the hot-plate. 
    External components can probe the temperature at this location at the current 
    simulation time. This necessitates the updating the heater state up-to the current time.
    If time at which the probing is performed is sooner than the next wake-up governed 
    by the time-step $\Delta t$ corresponding to the Forward Euler method, an update at 
    a smaller time-step is performed. Since this doesn't affect the stability of the 
    numerical method, the update can be safely performed. 
  \item {\bf Generated Events:} User can specify a threshold value of temperature 
    (or multiple values) at the probe location. Whenever the {\em{rising}} temperature at the probe 
    location crosses the threshold values, an event (probe-temperature crossed (rising)) 
    is generated. Similarly whenever the {\em{falling}} temperature at the probe location crosses the 
    threshold value, another event (probe-temperature crossed (falling)) is generated. 
    External processes waiting for these events to occur are then automatically notified  
    (using the {\tt yield <event>} construct of SimPy).
  \item {\bf Wake-up Events:} Since numerical solution of the governing PDE requires an 
    iterative time-stepping scheme, a wake-up event is generated after every time-step which 
    schedules the state-update for the next time-step. The time-step size is governed by the 
    stability requirements of the numerical method. 
    If any other event (probe/perturbation/generated) occurs before the scheduled wake-up takes 
    place, a state-update is scheduled by those events as described earlier. Once those events 
    are executed, the system resumes the wake-up cycle for advancing in time. 

\end{enumerate}

\subsubsection{Numerical Results}

\begin{figure}
\centering
\includegraphics[width=\textwidth]{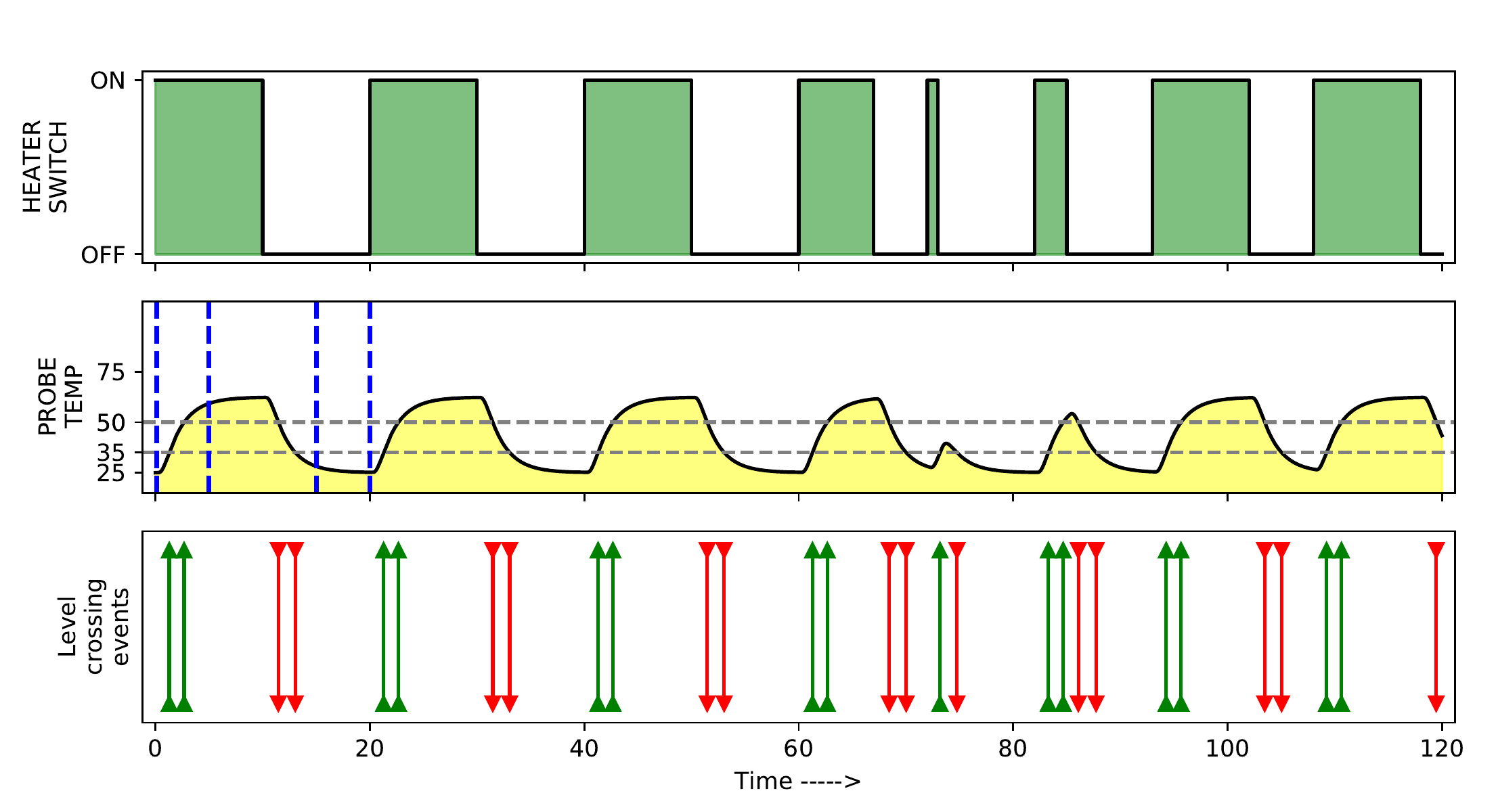}
\caption{Time evolution of the probe temperature showing events generated
  during simulation. The level crossing events (rising and falling) are shown with upward and 
  downward arrows respectively. The threshold temperatures were set at $35^0$C and $50^0$C.} \label{Fig:HeaterValid}
\end{figure}

\begin{figure}[h!]
    \begin{minipage}{0.5\textwidth}
        \centering
   		\includegraphics[scale=0.4]{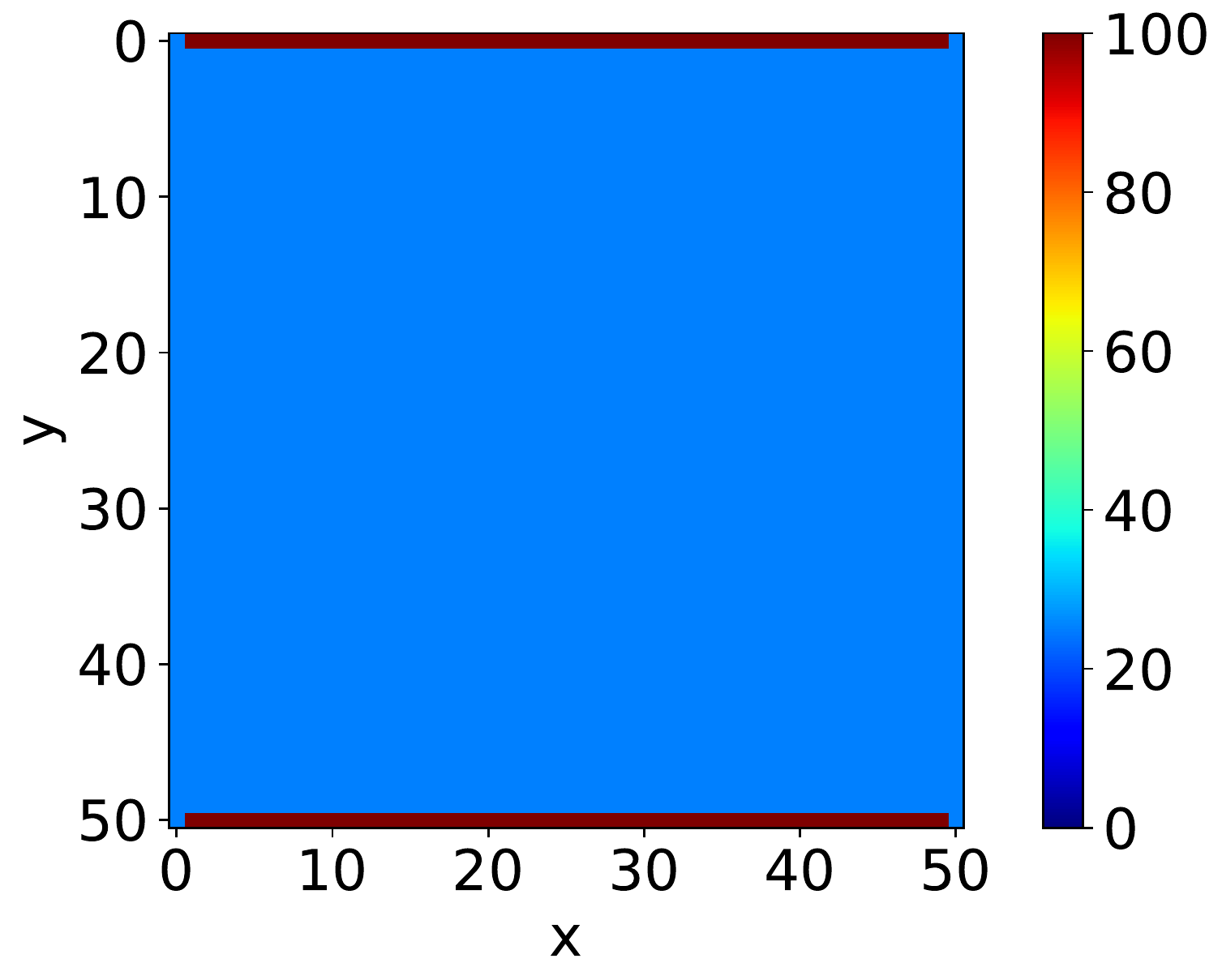}
      \subcaption{$T=\Delta T$min}
   \end{minipage}%
   \begin{minipage}{0.5\textwidth}
        \centering
   		\includegraphics[scale=0.4]{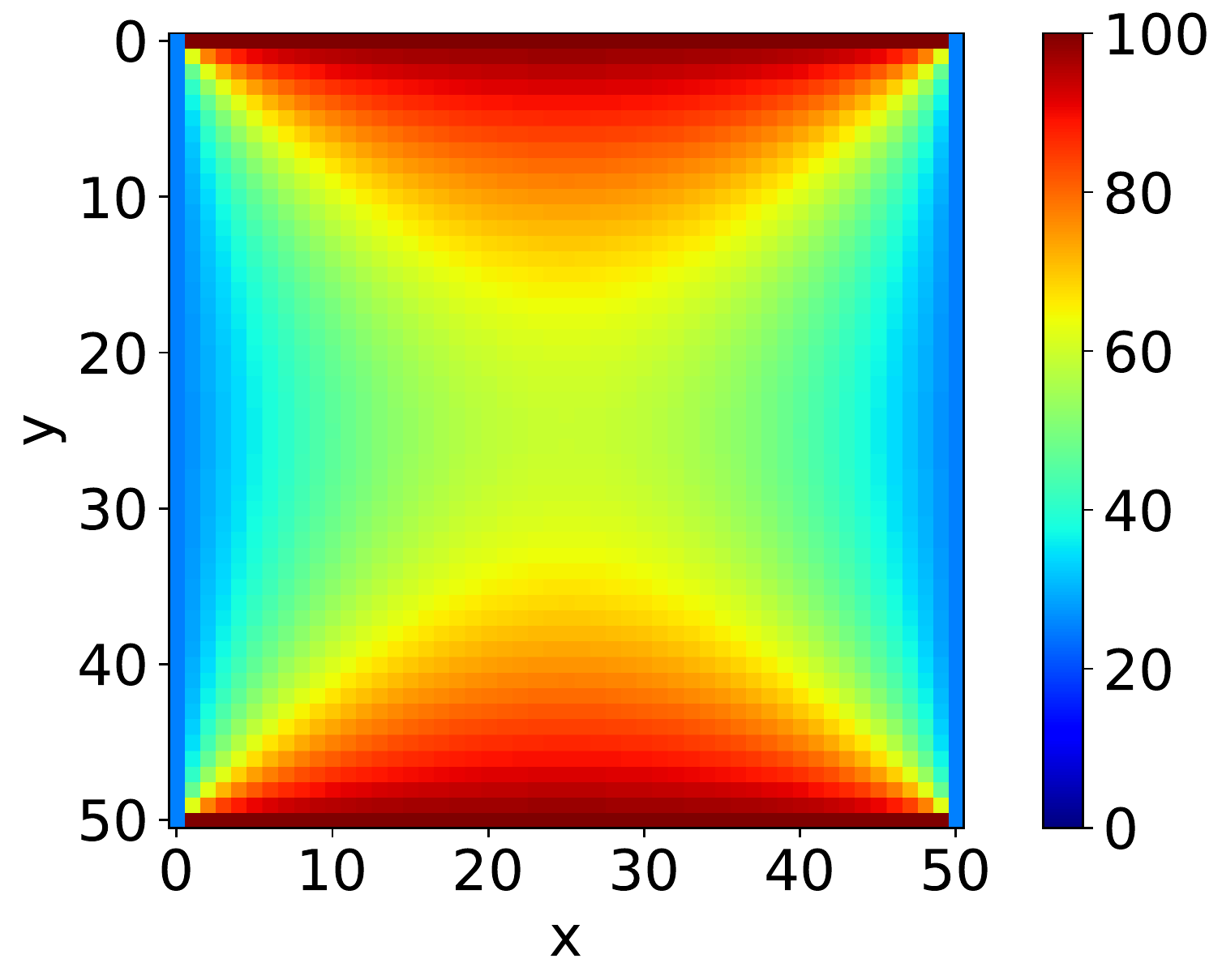} 
      \subcaption{$T=5$min}
   	\end{minipage}  \\
   	    \begin{minipage}{0.5\textwidth}
        \centering
   		\includegraphics[scale=0.4]{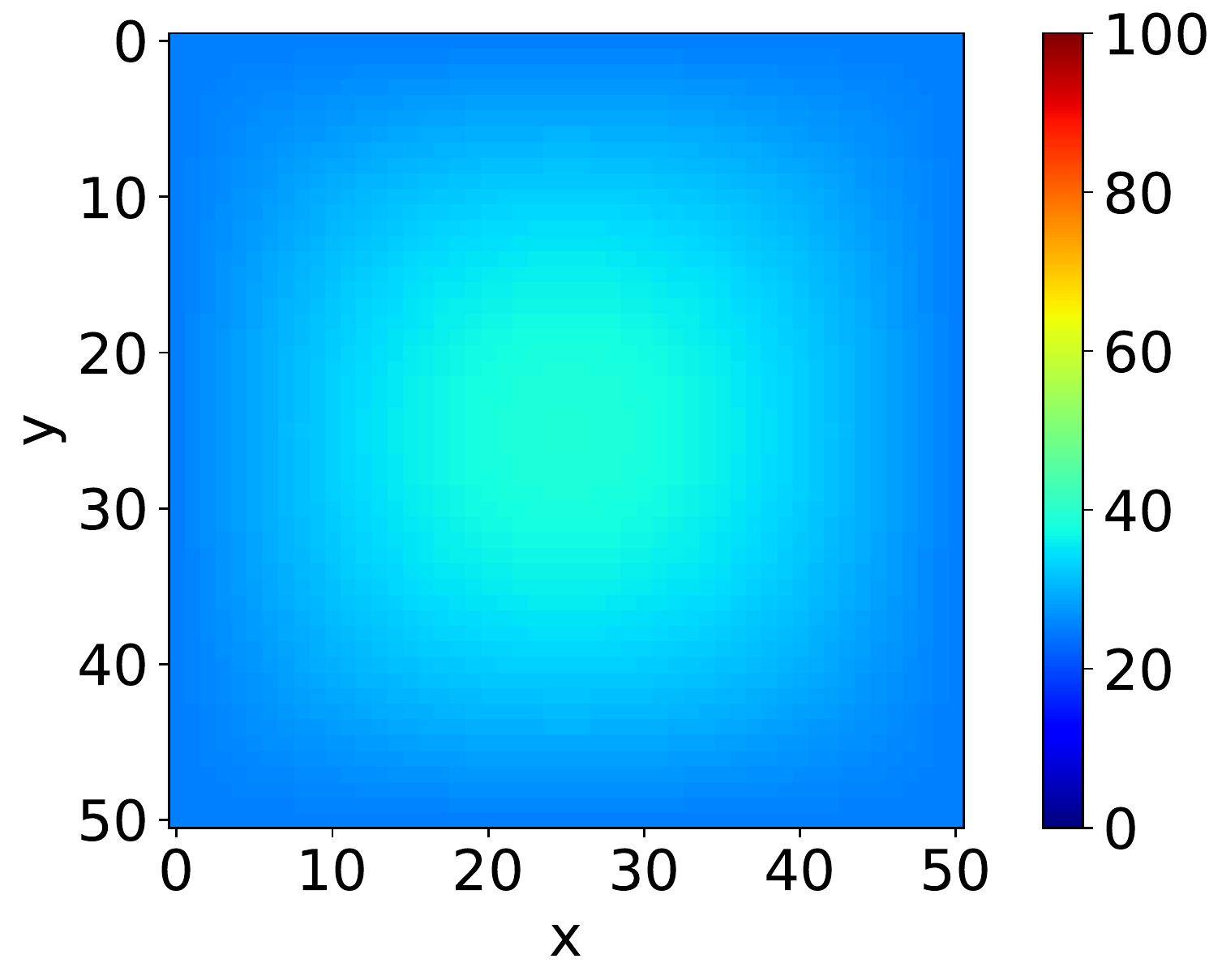}
      \subcaption{$T=15$min}
   \end{minipage}%
   \begin{minipage}{0.5\textwidth}
        \centering
   		\includegraphics[scale=0.4]{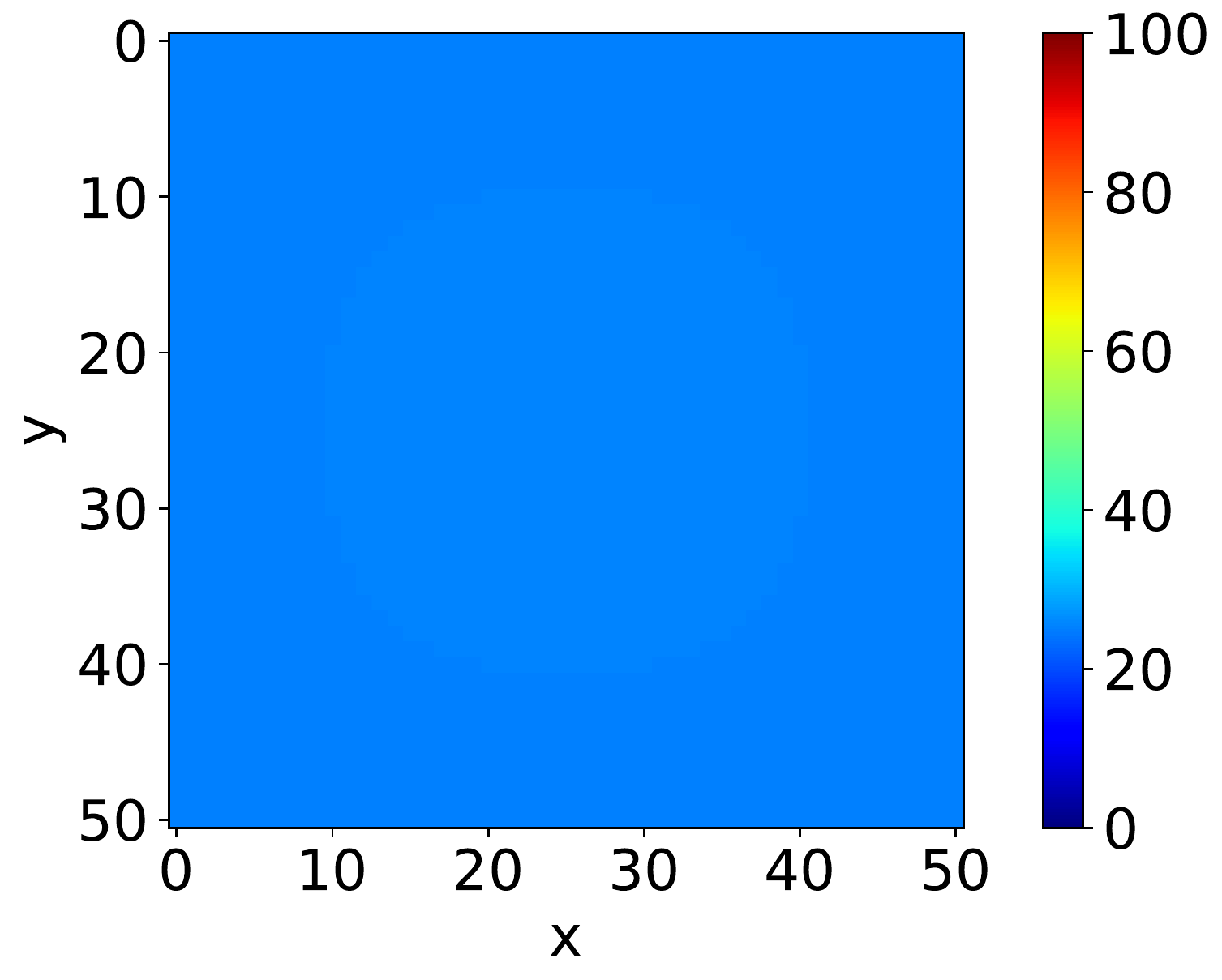}
      \subcaption{$T=20$min}
   	\end{minipage} 
    \caption{Temperature profile snapshots at time $T = \Delta t, 5, 15, 20$ mins.}
    \label{Fig:HeaterSnapshoptsValid}
\end{figure}

We validate the heater model as follows. 
A heating and cooling schedule is designed to validate the heater model. 
The schedule consists of two identical cycles of heating and cooling in the time span 
of first $50$min, and
a random heating and cooling schedule from $T=50$min to $T=120$min. 
Both cycles till $T=50$min are of identical time duration and consist of a heating process followed by
a cooling process. Figure \ref{Fig:HeaterValid} shows a time-series plot of the 
temperature at the center of the hot-plate (measured by the probe). 
The framework allows specifying threshold temperature values to schedule Generated Events. 
In this example, we have set
two temperature values, i.e. $u = 35^0$C and $u=50^0$C as shown in the figure. 
Whenever the {\em{rising}} temperature at the probe location crosses the threshold values, a Generated
Event (probe-temperature rising, denoted by green upward pointing arrow) is created. 
Similarly whenever the {\em{falling}} temperature crosses
the threshold values, another generated event
(probe-temperature falling, denoted by red downward pointing arrow) 
is created. These generated events can affect either the discrete event
schedule or other continuous systems in the framework.
The entire simulation is run through a discrete event scheduler, which schedules a 
wake-up event for each time step to advance the simulation in time. 
It can be seen that, the temperature at the center of the hot-plate starts increasing
when the heater is turned {\tt ON} 
and asymptotically reaches a steady-state value of $62.8^0$C. 
At time $T = 12.5$min, the heater is turned {\tt OFF} and the temperature starts dropping 
exponentially until it reaches the {\tt LOW\_temp} value of $25^0$C. 
As the heat loss takes place only at the edges, a higher residual temperature remains at the
interior parts of the hot-plate for a short time even after the heater is turned {\tt OFF}. 
Eventually the heat loss results in a uniform temperature of $25^0$C over the entire 
area of the hot-plate. 
Figure \ref{Fig:HeaterSnapshoptsValid} shows snapshots of the hot-plat at time
$\Delta t, 5, 15$ and $20$ minutes, 
where, $\Delta t$ is the time after the first time-step is performed.
Corresponding time-values are highlighted in \ref{Fig:HeaterValid} with dashed blue lines. 
It can be clearly seen that, at $\Delta t$ the two opposite edges are at {\tt HIGH\_temp}
while the rest of the hot-plate is at {\tt LOW\_temp}.
The value of $\Delta t$ is dictated by the stability conditions as stated earlier.
At $T = 5$min, heat has dissipated inside the domain raising the temperature
differentially at different parts. 
After the heating is turned {\tt OFF}, a residual high temperature is observed
for a short while in the interior parts of the plate, for example as shown 
in the figure at $T = 15$min. 
This residual temperature returns to {\tt LOW\_temp} as the heat loss takes place through 
all four boundaries. 
Sometime before $T = 20$min, the temperature over the entire domain asymptotically 
returns to $25^0$C. The cycle repeats after $T = 25$min till $T=50$min. 
After $T=50$min, the heating and cooling schedule is kept random. This is to demonstrate that
the heating and cooling can be randomly applied in the ongoing simulation and 
an a-priori knowledge of the same is not required for running the simulation.
This has been made possible due to our approach to time-stepping the simulation via a 
discrete event scheduler, including for the embedded continuous processes. 
In the subsequent sections, we demonstrate an example where the heater works along with 
another continuous simulation entity (water tank) in the discrete event framework, such 
that both the discrete event framework and the continuous entities can potentially have an
effect on each other.

\subsection{Fluid Tank}\label{sec:tank}

	\subsubsection{Description:}
	Consider a fluid tank whose level is to be simulated and monitored
	continuously with respect to time. The tank has inlet and outlet valves
	that can be opened and closed based on external triggers. Opening/closing
	of these valves changes the state trajectory of the tank. The flow rate
	through the inlet/outlet are assumed to be fixed parameters in the model.
	They are specified in units of length per-time and defined as maximum
	change in the fluid level per unit time when the corresponding valve is
	open. Further, the maximum level at which the tank is considered full, is
	also a parameter.

\begin{figure}
	\centering
	\includegraphics[width=\textwidth]{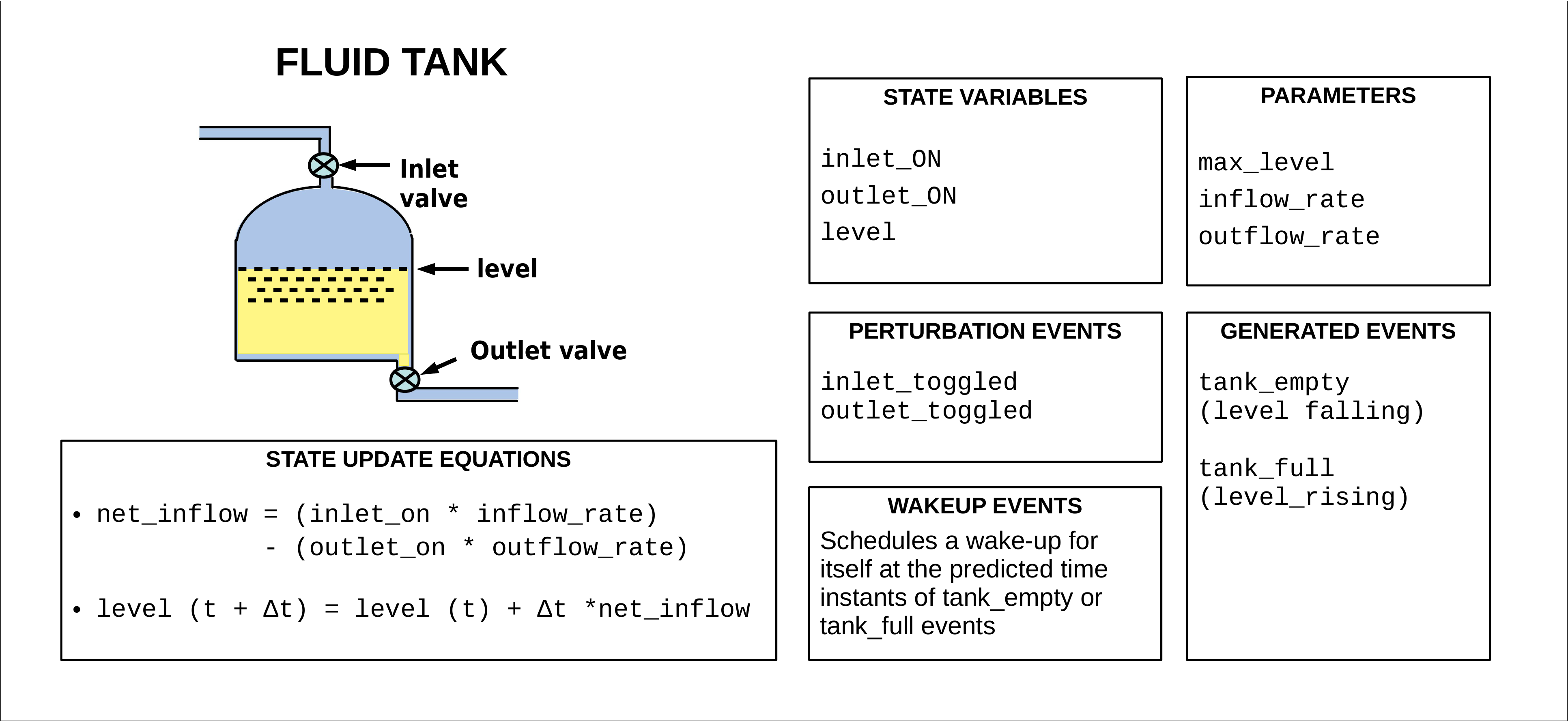}
	\caption{Model of a fluid tank} \label{fig:tank}
	\end{figure}

	The state-update equations for the tank, along with a summary of the state
	variables and parameters is presented in Figure \ref{fig:tank}. For this
	continuous entity, the state-update equation is a simple linear algebraic
	equation and therefore, an iterative time marching method is not necessary.
	The state after a given time interval $\Delta t$ can be directly computed
	from the current state as long as the condition of the inlet/outlet valves
	do not change.
	We now describe how this continuous entity interacts with the external
	processes, and how its simulation can be performed by time-advancement
	through the discrete-event engine.
	
	\subsubsection{Discrete-Event Interface and Simulation:}
	The tank entity can be implemented as a Python class with an interface
	similar to that in Figure \ref{fig:continuous_entity}. The state variables,
	parameters and the state-update function become members of this class.  The
	tank interacts with the environment via the following events serving as an
	interface:

	\begin{enumerate}
	
	\item {\bf Perturbation Events:} External events can cause the tank's inlet
	or outlet valves to toggle their state, which can change the trajectory of
	the level.
	
	\item {\bf Probe Events:} External components can probe the level of the
	tank at the current simulation time. This necessitates updating the tank
	state up-to the current time.
	
	\item {\bf Generated Events:} Whenever the tank level falls, and the tank
	becomes empty, a {\tt tank\_empty} event is generated and triggered by the
	tank entity itself.  External processes waiting for this event to occur are
	then automatically notified (using the {\tt yield <event>} construct of
	SimPy). Similarly, when the tank level is rising and reaches the maximum
	value, a {\tt tank\_full} event is generated.

	\item {\bf Wake-up Events:} Owing to the linear state-update equations, the
	state updates need not be performed periodically at fixed time-steps.
	Rather they can be performed directly at time-instants of interest.
	Whenever a state-update is performed, the future time instant at which the
	tank is expected to become empty (if the level is falling) or full (if the
	level is rising) is computed, and the entity schedules a wake-up event for
	itself at this precise time instant.  When the wake-up event (or any
	perturbation/probe/generated events) occur in the tank, the state-update is
	performed, empty/full events are triggered if the tank has become
	empty/full at this instant, and the next wake-up event is scheduled, based
	on the current trajectory. 
		
	If it so happens that tank's state trajectory changes sometime before the
	next wake-up event (for instance, due to the toggling of a valve), the
	state is updated and a new wake-up event is scheduled based on the updated
	trajectory. The old wake-up event however is not cancelled. It simply
	causes the tank state to be updated up-to the time instant of the old
	wake-up event and does not have any side effects on the state.
	\end{enumerate}

	\subsubsection{Validation and Simulation Results:}
\begin{figure}
	\centering
	\includegraphics[width=\textwidth]{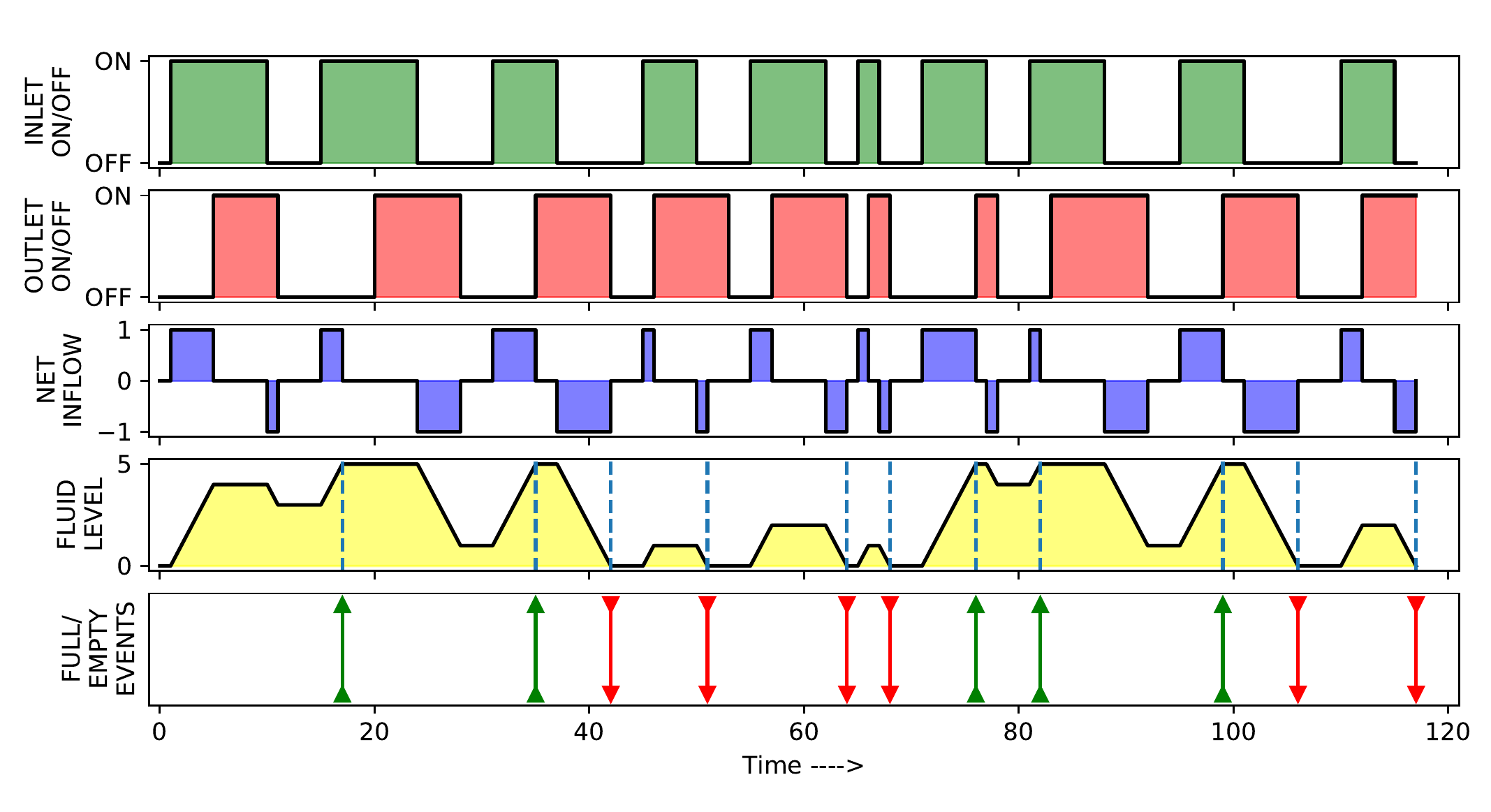}
	\caption{A plot of the time evolution of the tank's state showing the empty/full events generated during simulation.
	The inlet/outlet values are toggled after random time intervals} \label{fig:tankplots}
	\end{figure}

	Figure \ref{fig:tankplots} shows the simulation results obtained for 
	a simple validation exercise involving a single tank instance. Here, an external SimPy
	process toggles the tank's inlet and outlet valves after random time intervals.
	The generated {\tt tank\_empty} and {\tt tank\_full} events are also indicated in the plot.

\subsection{The System}\label{sec:System}

	We now describe the complete system where the fluid tank and heater
	instances interact with each other and with the environment 
	through the {\em events interface}.
	To illustrate a two-way dependency between the fluid tank and heater
	instances, we model the following interactions in the system:

	\begin{enumerate}
	
	\item At the start of simulation, the heater is assumed to be OFF 
	and the temperature of the entire heating plate is set to the {\tt LOW\_temp}
	value of $25^0$C.
	The fluid tank is assumed to be full. The tank's inlet
	and outlet valves are both closed.
	
	\item The heater is turned ON. As soon as the probe temperature crosses a
	certain threshold (in this case $50^0$C), the system can start
	processing external jobs one-by-one. The arrival of jobs is modeled by a stochastic
	discrete-event process.
	
	\item Each arriving job has a duration which is random and uniformly
	distributed between 0.5 to 1 minutes. Also the time interval between
	arrival of successive jobs is also uniformly distributed between 0.5 to 1
	minutes. The tank outlet valve needs to be kept open for the duration of
	each job. Thus the tank gradually empties as successive jobs are processed.

	\item As soon as the tank becomes empty, the heater is turned OFF, and the
	tank refill process is initiated.  The tank inlet valve is opened and the
	outlet valve is closed. The tank gradually becomes full again.

	\item As soon as the tank becomes full, the entire loop is repeated,
	starting from step 2.

	\end{enumerate}
	
\begin{figure}[t]
	\centering
	\includegraphics[width=\textwidth]{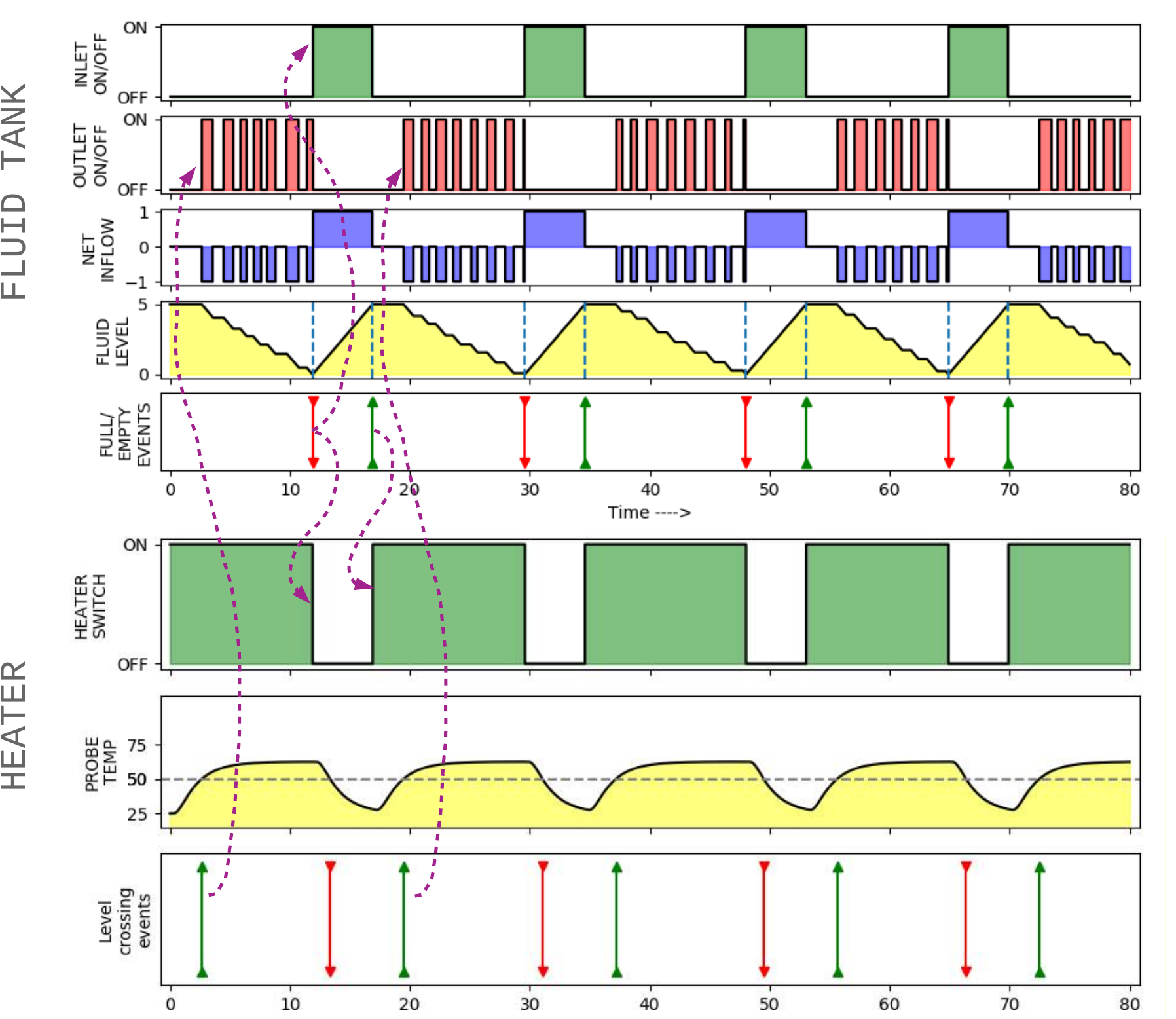}
	\caption{Simulation results for a system containing interacting tank and heater instances. The causality of events
	is indicated by dashed arrows.} \label{fig:systemplot}
	\end{figure}

	To implement these interactions, an additional Python class ({\tt Controller})
	is created. The behavior of the controller can be described in a straightforward
	manner as a SimPy process. Listing \ref{lst:Controller} is an excerpt from the 
	controller's behavioral loop described above. The code listing serves to highlight the ease with 
	which complex interactions between continuous and discrete-event entities can be 
	described by the modeler. Figure \ref{fig:systemplot} presents simulation results
	generated for this system. The causality of events is indicated in the figure 
	by dashed arrows. The simulation also produces a detailed event log.

\begin{lstlisting}[float, caption={Python excerpt for implementing the system control loop described in Section \ref{sec:System}}, label={lst:Controller}, language=SimPy]
# Python code excerpt describing the control loop
# note: env is an instance of SimPy's environment object

class Controller():
	# initialization, and other member functions
	# .....

	def controller_behavior(self,env):
		
		# initially the tank is full and its valves are closed.
		tank.level=tank.max_level
		tank.inlet_ON=False
		tank.outlet_ON=False
		# initially the heater is off. check this.
		assert(not heater.is_heater_ON())
		
		while(True):    
		
			# turn ON the heater 
			heater.turn_ON_heater_switch()
			# wait until the probe temperature goes 
			# above a set threshold.
			if(heater.probe_temp() < 50):
				yield heater.threshold_crossed_rising_event

			# now, start processing jobs by periodically
			# toggling the tank outlet valves. Do this until 
			# the tank level falls below a threshold.
			
			while(not tank.is_empty()):
				# turn ON the tank valve
				tank.toggle_outlet_valve()
				job_duration = random.uniform(0.5, 1)
				job_done   = env.timeout(job_duration)
				tank_empty = tank.empty_event

				# wait until the job is done or the tank becomes empty
				what_occured = yield (job_done | tank_empty)

				if job_done in what_occured: 
					# wait for some time and then repeat with next job
					tank.turn_OFF_outlet_valve()
					idle_duration = random.uniform(0.5, 1)
					yield env.timeout(idle_duration)
				else:
					# tank became empty.
					break
			# tank became empty. Now turn OFF the heater, 
			# and start the tank refilling process.
			# then, wait until the tank becomes full again.
			heater.turn_OFF_heater_switch() 
			tank.turn_OFF_outlet_valve()  
			tank.turn_ON_inlet_valve()
			yield tank.full_event

			# now the tank has become full.
			# turn off the tank valves.
			tank.turn_OFF_inlet_valve()
			tank.turn_OFF_outlet_valve()
			# done. repeat the loop.

\end{lstlisting}

\section{An Improved Scheme for Time Advancement}\label{sec:timeAdvancement}

As discussed in Section \ref{sec:framework}, time advancement can be performed
using either a predictive time-stepping approach, or using fixed time-steps
when the governing equations of the continuous entity require iterative state
updates.
One issue with the fixed time-step approach is that if the time step size for the continuous
simulation is very small relative to the typical time between events in the
rest of the system, the cost of adding a wakeup event to the global event list
after every time-step can be prohibitive.  To address this, the following modification can be used:
	
	At each iteration ($K$) of the event-stepped algorithm, the {\em tentative}
	time-step ($\Delta_K$) is taken to be the difference between the scheduled
	time of the next event in the global event list ($t_\textrm{next\_event}$)
	and the simulation time for the current iteration($t_{K}$). 
	$$\Delta_K = t_\textrm{next\_event} - t_\textrm{K}$$
	Each continuous entity is then asked to {\em peek-ahead} in time by a total
	period of $\Delta_K$ by executing its state-update equations. This can be
	done either using a single step of size $\Delta_K$ or by dividing this
	period into finer time-steps as as dictated by the time marching scheme.
	The computation of state updates for multiple continuous entities can
	potentially be executed in parallel.
	If no output events of interest are predicted to be generated by any of the
	continuous entities in this period, the time can be advanced by $\Delta_K$
	and the computed state-updates in each of the continuous entities can be
	applied before proceeding to the next iteration.  However, if it is found
	that for a continuous entity $i$, an event of interest is generated at time
	$t_i < t_\textrm{next\_event}$, then it may be possible that this event could
	affect the state or trajectories of the other entities in the system. Thus
	the actual time-step taken must be the one that advances time to the
	earliest predicted event across all of the continuous entities.  That is,
	the simulation time should be advanced to $t_{K+1} = \textrm{min}_i(t_i)$ in
	the next iteration.

	The earliest event predicted to occur at time $t_{K+1}$ can then be
	inserted into the global event-list, so that its effect on other entities
	can be propagated as usual in a discrete-event framework, and the state
	updates in all of the continuous entities computed up to time $t_{K+1}$ can
	be applied before advancing simulation time to $t_{K+1}$.  A further
	optimization is to adaptively adjust the tentative time step $\Delta_K$ for
	improved performance.
	Implementing this requires the continuous simulation framework to support 
	a peek-ahead or roll-back feature.

	A second approach is to use meta-modeling (in the form of regression-based models, 
	neural networks with supervised learning, or physics-informed neural networks)
	for predicting the time instants of generated events in advance. 
	In fact, detailed models of continuous processes can often be replaced by
	reduced order surrogate models when accuracy needs to be traded for evaluation speed.
	In such models, the time instants of generated events can be predicted ahead of time, 
	and the predictive time-stepping approach can be used for efficient simulation. 
	Exploring the use of these approaches for building digital twins is a promising direction.

\section{Future Work and Conclusions}\label{sec:futureWork}
	
	In this paper we presented a Python based Mixed Discrete-Continuous Simulation (MDCS)
	framework specifically targeted for Digital Twins applications. The
	framework is based on Python's SimPy library and uses its event-stepped 
	algorithm for coordinating the time advancement. We presented a detailed example 
	of interacting continuous entities simulated using this framework.
	The design aspects of the framework and the simulation approach make it
	well-suited for digital twin applications for the following reasons: 
		
		\begin{itemize}
		
		\item The event-stepped approach can result in a more
		efficient simulation for scenarios where only a few kinds of events affect the
		trajectory of continuous entities in the system. 
		
		\item In this approach, it is
		possible for different continuous entities in the system to use different
		continuous solvers and internal time step-size values. 
		
		\item The loose coupling between the continuous entities presents opportunities for
		executing their behavior in parallel within a single time-step for real-time simulation. 
		
		\item For modeling
		entities where a high level of accuracy may not be necessary, coarse surrogate
		models can be used to predict the trajectory and time of generated events and
		schedule a wakeup ahead of time.  
		
		\item Sensor value updates from the real
		system can be easily incorporated into the simulation as perturbation events
		affecting the state.  
		
		\end{itemize}

	Future improvements to the framework are planned along the following directions:
	\begin{enumerate}
	\item \textbf{Integration with existing continuous simulation frameworks}

	For fast simulation of continuous processes, integration with established continuous
	simulation frameworks is necessary. This requires building wrappers for invoking 
	state update functions of continuous solvers such as OpenFOAM. We also plan to integrate
	Dolfin \cite{dolfin10}, a Python based finite-element library for multiphysics 
	modeling and simulation. 

	\item \textbf{Incorporating analytics}

	For Digital Twin applications, analytics modules need to be incorporated into the 
	framework for
	parameter extraction from sensor data, prediction, optimization and for building 
	surrogate models in run-time. 

	\item \textbf{Acceleration for real-time simulations}

	The requirement for real-time simulation creates a need for simulation acceleration 
	that is possible using hardware platforms such as GPGPUs,
	FPGAs or parallel execution on multi-core systems. 
	It is possible to explore architectures that can take advantage of these technologies
	for simulations.

	\item \textbf{Support for sensing and control}

	Sensing and control are integral aspects of a Digital Twin. Features that support 
	these aspects need to be explored and integrated into the framework. 
	\end{enumerate}


%
\bibliographystyle{splncs04}
\bibliography{references}
\end{document}